\documentclass[%
 aip,
 amsmath,amssymb,
 reprint,%
]{revtex4-1}

\usepackage{graphicx}
\usepackage{dcolumn}
\usepackage{bm}

\usepackage[utf8]{inputenc}
\usepackage[T1]{fontenc}
\usepackage{mathptmx}
\usepackage{etoolbox}
\usepackage{bm}
\usepackage{graphviz}
\usepackage[autosize]{dot2texi}
\usepackage{tikz}
\usepackage{float}
\DeclareMathOperator*{\argmin}{argmin}

\makeatletter
\def\@email#1#2{%
 \endgroup
 \patchcmd{\titleblock@produce}
  {\frontmatter@RRAPformat}
  {\frontmatter@RRAPformat{\produce@RRAP{*#1\href{mailto:#2}{#2}}}\frontmatter@RRAPformat}
  {}{}
}%
\makeatother
\begin{document}

\preprint{AIP/123-QED}

\title[]{Gradient enhanced multi-fidelity regression with neural networks:\\ application to turbulent flow reconstruction}
\author{Mohammad Hossein Saadat}
 \email{mohammad.saadat@alumni.ethz.ch}
 \affiliation{{Department of Mechanical and Process Engineering, ETH Zurich, 8092 Zurich, Switzerland}}


\date{\today}

\begin{abstract}
A multi-fidelity regression model is proposed for combining multiple datasets with different fidelities, particularly abundant low-fidelity data and scarce high-fidelity observations. The model builds upon recent multi-fidelity frameworks based on neural networks, which employ two distinct networks for learning low- and high-fidelity data, and extends them by feeding the gradients information of low-fidelity data into the second network, while the gradients are computed using automatic differentiation with minimal computational overhead. The accuracy of the proposed framework is demonstrated through a variety of benchmark examples, and it is shown that the proposed model performs better than conventional multi-fidelity neural network models that do not use gradient information. Additionally, the proposed model is applied to the challenging case of turbulent flow reconstruction. In particular, we study the effectiveness of the model in reconstructing the instantaneous velocity field of the decaying of homogeneous isotropic turbulence given low-resolution/low-fidelity data as well as small amount of high-resolution/high-fidelity data. The results indicate that the proposed model is able to reconstruct turbulent field and capture small scale structures with good accuracy, making it suitable for more practical applications. 
\end{abstract}

\maketitle

\section{Introduction} \label{}
We are currently living in the era of "big data", yet acquiring accurate high-fidelity data, either through high-precision experimental measurements or high-resolution numerical simulation, is still hard and expensive due to limited experimental or computational resources. Low-fidelity data, on the other hand, are more accessible as they require fewer resources and are not as computationally demanding, often relying on reduced-order-models or coarse numerical simulations.
While low-fidelity data may not be as precise as high-fidelity data, they can still capture the overall trend and are usually strongly correlated to their high-fidelity counterparts \cite{meng2020composite}.
The key idea of multi-fidelity modeling is, then, to utilize data-driven techniques to leverage abundant low-fidelity data, combine and blend it with scarce high-fidelity data, exploit the cross-correlation between them, and ultimately achieve a model that outperforms a model solely rely on few high-fidelity data. 

Many data-driven techniques have been proposed in the literature so far for multi-fidelity modeling \cite{perdikaris2017nonlinear, parussini2017multi, lee2019linking, deng2020multifidelity, stanek2021multifidelity}. In recent years, however, data-driven methods based on deep neural networks (DNNs) have become increasingly popular in the computational science community owing to the huge success of deep learning in doing difficult tasks, from object detection to speech recognition and natural language processing. DNNs have been used to solve, possibly ill-posed, partial differential equations \cite{raissi2019physics, eivazi2022physics}, identify governing equations of complex nonlinear dynamics \cite{raissi2018hidden, chen2021physics} and construct reduced-order-models of high-dimensional systems \cite{fukami2021model}, to name a few. DNNs have also recently received attention for multi-fidelity modeling \cite{meng2020composite, guo2022multi, conti2023multi}, as their inherent non-linearity allows them to learn complex non-linear correlation between low- and high-fidelity data. In particular, the NN architecture proposed in \cite{meng2020composite, guo2022multi} showed promising results and inspired several 
other multi-fidelity models for different applications \cite{conti2023multi, partin2023multifidelity, pawar2022multi, meng2021multi}. 

Building on the works in \citep{meng2020composite, guo2022multi}, this paper extends the "2-step" NN model presented in \cite{guo2022multi} by incorporating the gradient information of low-fidelity data into the network. The proposed model consists of two distinct neural network: one for learning the low-fidelity data and the other one for learning the cross-correlation between low- and high-fidelity data, with the gradient of low-fidelity data being fed into the second network as an additional variable.
This incorporation of gradient information is crucial as, we shall see below, it improves the accuracy and performance of the multi-fidelity model \cite{perdikaris2017nonlinear, lee2019linking, deng2020multifidelity}. The computation of derivatives is done using automatic differentiation (AD) technique through back-propagation, which is already available in most of the popular deep learning packages, such as PyTorch or TensorFlow. It is shown that the procedure outlined above, results in a multi-fidelity model that performs better than "2-step" NN models that do not use gradient information, while maintaining a comparable level of computational cost and complexity. Furthermore, motivated by recent studies on the use of periodic periodic functions as an alternative to traditional activation functions \cite{sopena1999neural, parascandolo2016taming, sitzmann2020implicit, wong2022learning}, and their ability in representing complex physical signals \cite{sitzmann2020implicit}, the $\sin$ activation function is employed in this study. 

The other contribution of this work is to investigate the effectiveness of multi-fidelity modeling in turbulent flow reconstruction by focusing on the challenging case of decaying of the homogeneous isotropic turbulence. In particular, the reconstruction of instantaneous velocity field is studied given coarse simulation data as well as limited amount of high-resolution/high-fidelity data. It is demonstrated that the model is able to reconstruct turbulent field with good detail. 

The remainder of the paper is organized as follows: Section~\ref{Sec:Method} outlines the main steps of the proposed NN model for multi-fidelity regression. Section~\ref{Sec:Validation}, evaluates the accuracy and effectiveness of the proposed model for various benchmark test-cases. Application of the proposed model in turbulent flow reconstruction is studied in Section~\ref{Sec:Turb}, focusing on the decaying of the homogeneous isotropic turbulence. Finally, Section~\ref{Sec:conclusion} summarizes the results and provides conclusions.

\section{Methodology} \label{Sec:Method}
Let us consider the low-fidelity dataset $\mathcal{D}_{LF} = \left(\bm{x}_{LF}^{(i)}, y_{LF}^{(i)} \right)_{i=1}^{i=N_{LF}}$ , where $\bm{x}_{LF}^{(i)} \in \mathcal{R}^{d}$ are some observed locations at which we have the output of the inexpensive low-fidelity model $y_{LF}^{(i)}$ (for simplicity we assume $y_{LF}^{(i)} \in \mathcal{R}$). The high-fidelity dataset is defined in a similar way as $\mathcal{D}_{HF} = \left(\bm{x}_{HF}^{(i)}, y_{HF}^{(i)} \right)_{i=1}^{i=N_{HF}}$, where $y_{LF}^{(i)}$ denote the output of the costly but accurate high-fidelity model, and $N_{LF}$, $N_{HF}$ represent the number of low- and high-fidelity samples, respectively. The goal of multi-fidelity modeling is to discover the relation between low- and high-fidelity data \cite{meng2020composite},
\begin{align}
    y_{HF} = \mathcal{F}\left( \bm{x}, y_{LF} \right). 
    \label{eq:Discrete_Boltzmann_Eq}
\end{align}
Here, we approximate the unknown function $\mathcal{F}(.)$ with a deep neural network of the form,
\begin{align}
    y_{HF} \approx \mathcal{NN}_{HF} \left(\bm{x}_{HF}, y_{LF}\big|_{\bm{x}_{HF}}, \nabla_{\bm{x}} y_{LF}\big|_{\bm{x}_{HF}}; \bm{\theta}_{HF} \right),
    \label{eq:Discrete_Boltzmann_Eq}
\end{align}
where $\bm{x}_{HF}$, $y_{LF}\big|_{\bm{x}_{HF}}$ and $\nabla_{\bm{x}} y_{LF}\big|_{\bm{x}_{HF}}$ are inputs to the network $\mathcal{NN}_{HF}$, and $\bm{\theta}_{HF}$ is a collection of all unknown network parameters including weights and biases. In this network, the low-fidelity observation $y_{LF}$ and its gradient $\nabla_{\bm{x}} y_{LF}$ evaluated at high-fidelity locations $\bm{x}_{HF}$, are not readily available. Consequently, we first need to use another neural network $\mathcal{NN}_{LF}$,
\begin{align}
     y_{LF} \approx \mathcal{NN}_{LF} \left(\bm{x}_{LF}; \bm{\theta}_{LF} \right),
\end{align}
to approximate $y_{LF}$, from which the gradient $\nabla_{\bm{x}} y_{LF}$ can easily be computed through automatic differentiation.
\begin{figure*}
    \centering
    \includegraphics[width=1.0\textwidth]{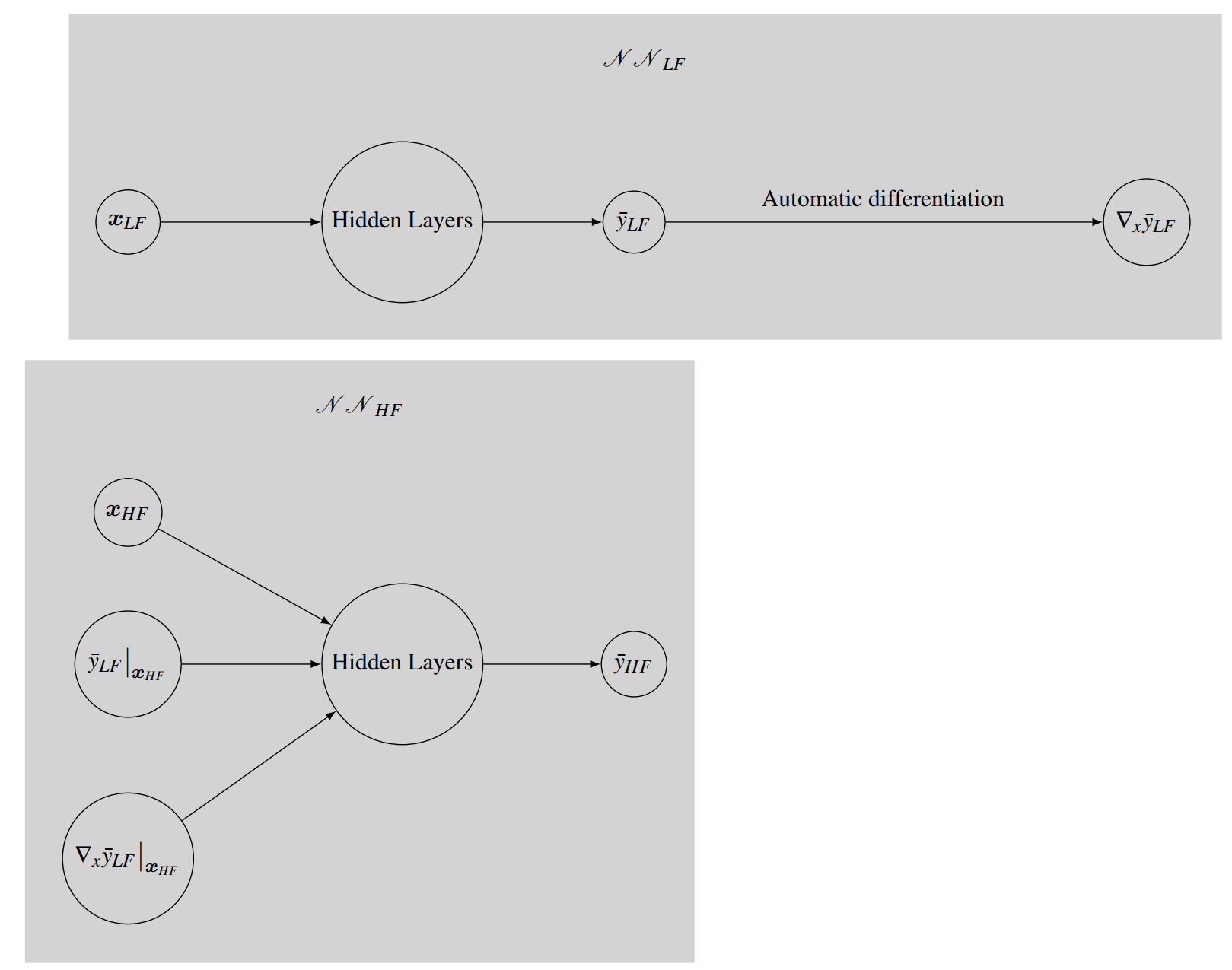}
    \caption{Schematic representation of proposed gradient enhanced multi-fidelity model (GH-MFR). The first network $\mathcal{NN}_{LF}$ reconstructs the low-fidelity data and its gradient, while the second network $\mathcal{NN}_{HF}$ uncovers the relation between low- and high-fidelity data and reconstructs the latter.}
    \label{fig:MFRmodel}
\end{figure*}

The parameters $\bm{\theta}_{LF}$ of the first network $\mathcal{NN}_{LF}$ can be learned by minimizing the following mean-squared loss function given by, 
\begin{align}
    \argmin_{\bm{\theta}_{LF}} \mathcal{L}_{LF} &= \frac{1}{N_{LF}} \sum_{i=1}^{N_{LF}} \|   \bar{y}_{LF}^{(i)} - y_{LF}^{(i)} \|^2,
    \label{eq:losslf}
\end{align}
while the loss function of the second network $\mathcal{NN}_{HF}$ can similarly be written as,
\begin{align}
    \argmin_{\bm{\theta}_{HF}} \mathcal{L}_{HF} &= \frac{1}{N_{HF}} \sum_{i=1}^{N_{HF}} \|   \bar{y}_{HF}^{(i)} - y_{HF}^{(i)} \|^2 + \lambda \sum_{i=1}^{N_{HF}} \lvert{\bm{\theta}_{HF}^{(i)}}\rvert,
    \label{eq:losshf}
\end{align}
where $\bar{y}_{LF}$ and $ \bar{y}_{HF}$ denote the outputs of the $\mathcal{NN}_{LF}$ and $\mathcal{NN}_{HF}$, and $\lambda$ is the $L_1$ regularization coefficient which is typically used to prevent overfitting. Figure.~\ref{fig:MFRmodel} illustrates the schematic of the gradient enhanced multi-fidelity model explained above.

\section{Validation}\label{Sec:Validation}
In this section, we study the accuracy and performance of the proposed multi-fidelity model on a variety of benchmark problems from the literature \cite{meng2020composite, perdikaris2017nonlinear, guo2022multi}. 

For all numerical experiments \textit{Adam} optimization algorithm is used to minimize loss functions and the network is initialized using the \textit{Xavier} scheme. The network architecture used at both steps consists of $2$ hidden layers with $200$ neurons in each layer and $\sin$ activation. All inputs to the networks are normalized between $0$ and $1$. 

\subsection{Continuous function with linear correlation}
We first examine a case with linear correlation between low- and high-fidelity data generated from,
\begin{align*}
    y_{LF} &= 0.5(6x - 2)^2\sin (12x-4) + 10(x-0.5) + 5,  \\
    y_{HF} &= (6x - 2)^2\sin (12x-4).
\end{align*}
The low-fidelity data $x_{LF}$ are sampled at $11$ equally spaced values in $x\in[0,1]$, while the high-fidelity data are given at $x_{HF}=\{ 0, 0.4, 0.6, 1\}$. The learning rate and regularization rate $\lambda$ are $0.001$.

\begin{figure}
    \centering
     \includegraphics[width=0.5\textwidth]{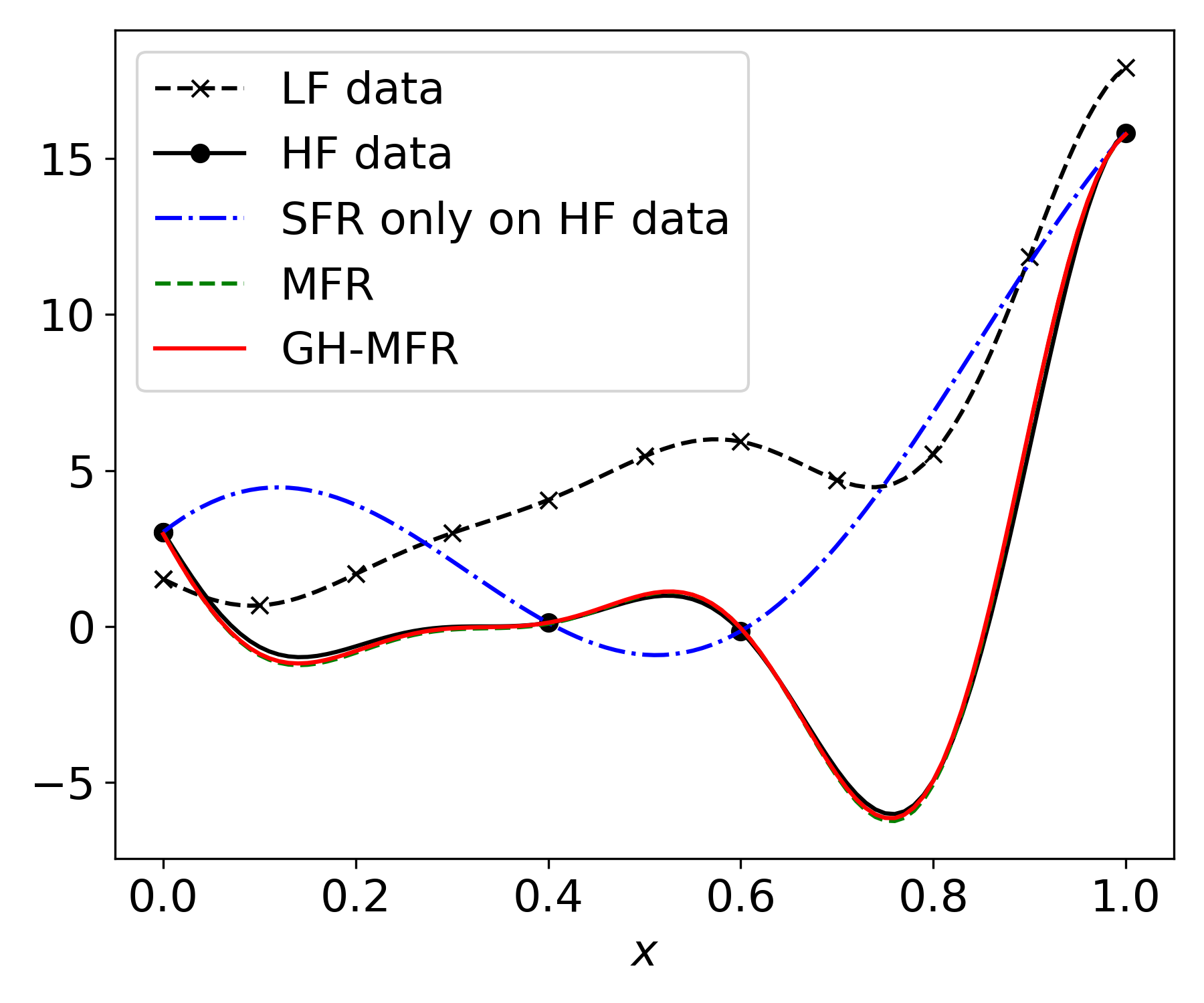}
    \caption{Results of gradient-enhanced multi-fidelity regression (GH-MFR) for case A with $11$ low-fidelity (LF) and $4$ high-fidelity (HF) observations. Performance is compared to multi-fidelity regression without gradient information (MFR) and single-fidelity regression trained on HF data only (SFR).}
    \label{fig:A}
\end{figure}

Figure~\ref{fig:A} shows the results of the present gradient enhanced multi-fidelity regression (GH-MFR) in comparison to the multi-fidelity model model without using gradient information (MFR) and also a single-fidelity model (SFR) trained only on high-fidelity data. It is evident that, the network trained only on high-fidelity data $x_{HF}$ (i.e., SFR) fails to correctly recover the exact high-fidelity data, while both multi-fidelity models give accurate results. 

\subsection{discontinuous function with linear correlation}
The second example studies the accuracy of the proposed model in cases with discontinuities in the solution. Here, the low- and high-fidelity data are generated according to equations given,
\begin{align*}
  y_{LF} &=
    \begin{cases}
      0.5(6x - 2)^2\sin (12x-4) + 10x - 10  \text{ if $0\leq x \leq 0.5$},\\
      0.5(6x - 2)^2\sin (12x-4) + 10x - 7  \text{ if $0.5\leq x \leq 1$},
    \end{cases}       \\
    y_{HF} &= 
    \begin{cases}
      2 y_{LF}(x) - 20x + 20 & \text{if $0\leq x \leq 0.5$},\\
      2 y_{LF}(x) - 20x + 24& \text{if $0.5\leq x \leq 1$},
    \end{cases}       
\end{align*}
\begin{figure}
    \centering
     \includegraphics[width=0.5\textwidth]{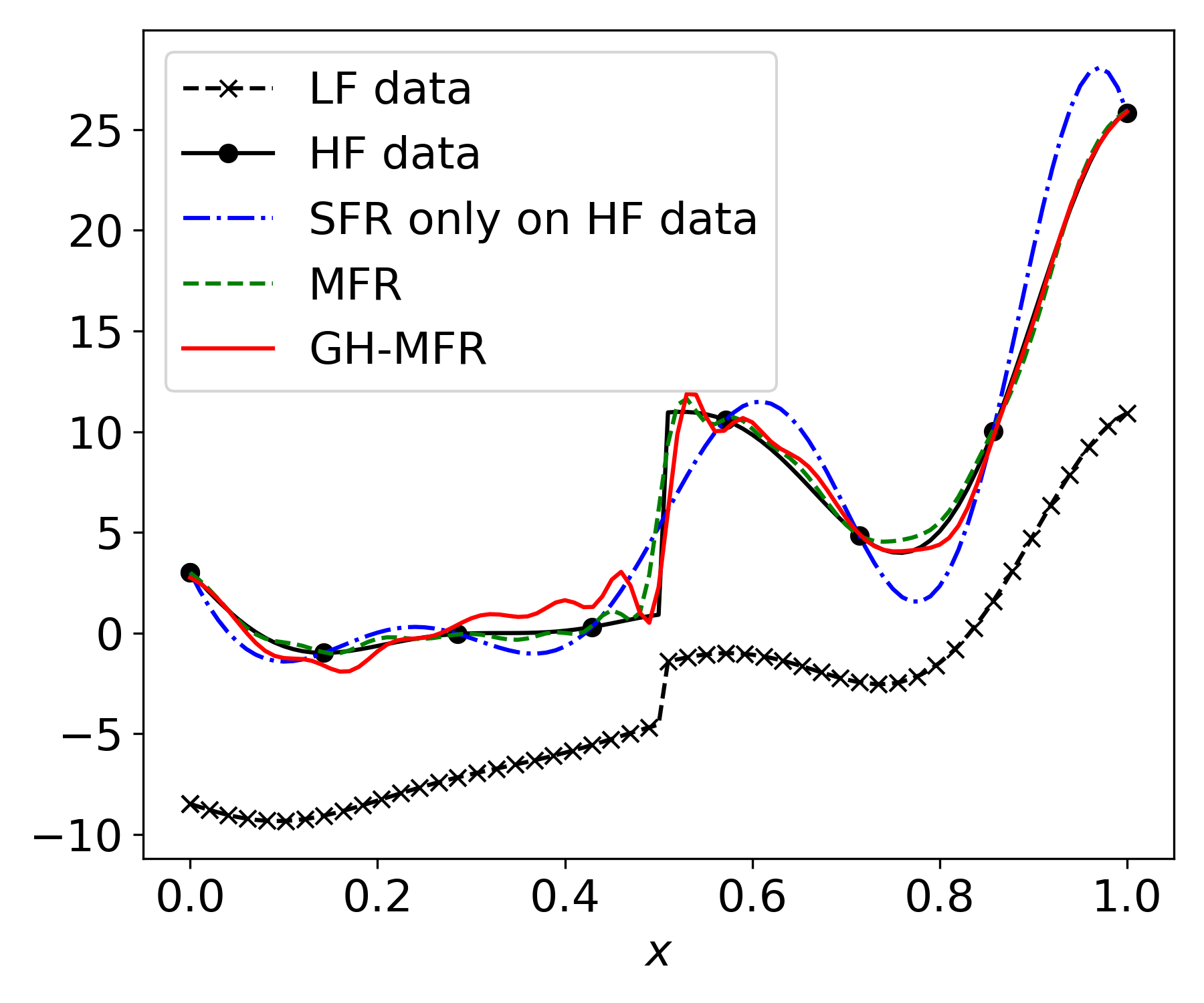}
    \caption{Results of gradient-enhanced multi-fidelity regression (GH-MFR) for case B with $50$ low-fidelity (LF) and $8$ high-fidelity (HF) observations. Performance is compared to multi-fidelity regression without gradient information (MFR) and single-fidelity regression trained on HF data only (SFR).}
    \label{fig:B}
\end{figure}
and $50$ and $8$ equally distanced points are sampled for the respective data and the regularization rate is set to $\lambda=0.001$. The results obtained by present multi-fidelity regression model is shown in Fig.~\ref{fig:B}. It is observed that the model is able to handle discontinuity, although small spurious oscillations appear around discontinuity. The single-fidelity model trained on high-fidelity data (SFR), however, is unable to capture the discontinuity and provide accurate results.

\subsection{Continuous function with nonlinear correlation}
The accuracy of the model in capturing non-linear correlations is assessed by considering a case with low- and high-fidelity data sampled from,
\begin{align*}
    y_{LF} &= \sin (8 \pi x),  \\
    y_{HF} &= (x - \sqrt{2}) \sin^2 (8 \pi x).
\end{align*}
\begin{figure}
    \centering
    \includegraphics[width=0.5\textwidth]{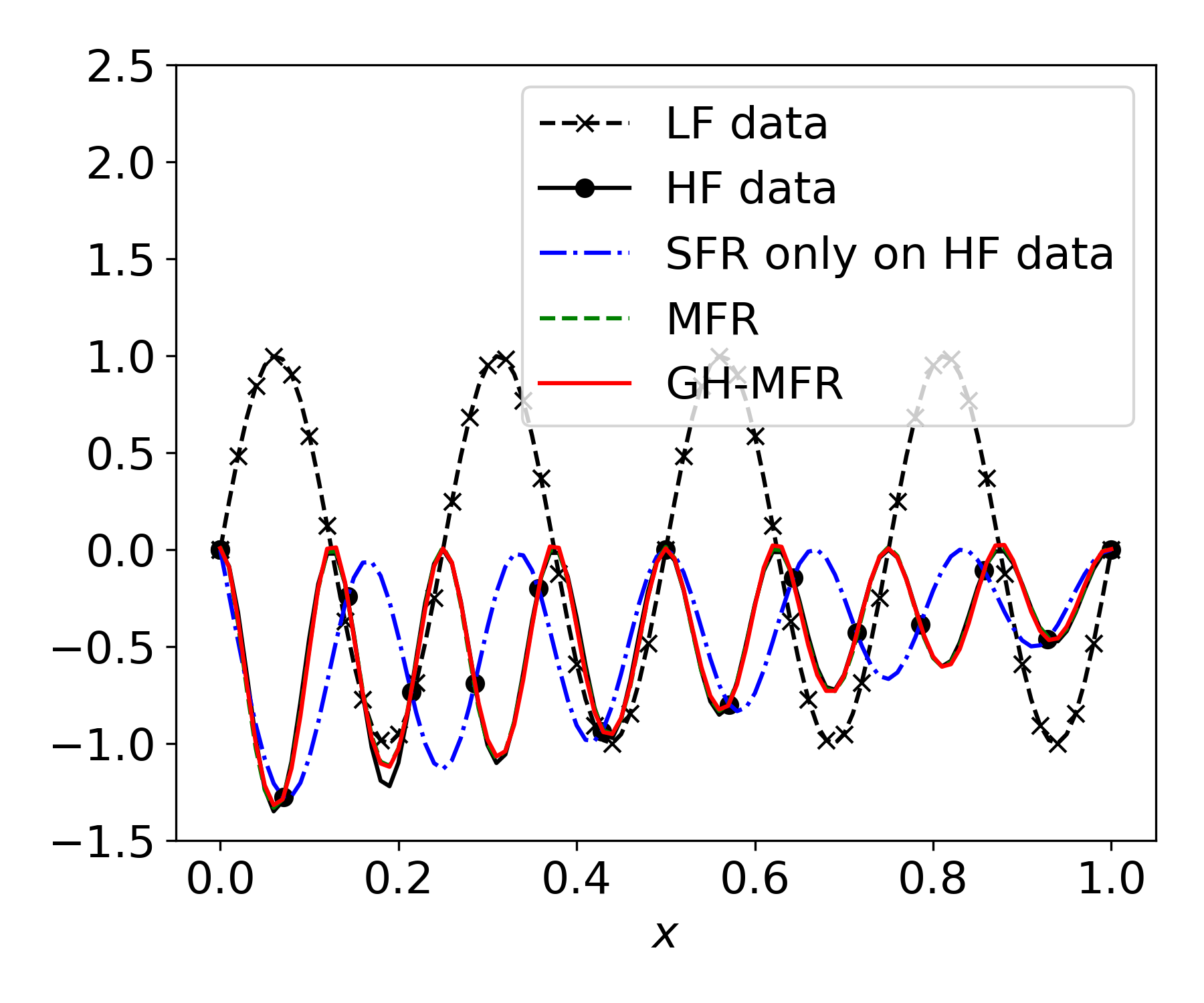}
    \caption{Results of gradient-enhanced multi-fidelity regression (GH-MFR) for case C with $51$ low-fidelity (LF) and $15$ high-fidelity (HF) observations. Performance is compared to multi-fidelity regression without gradient information (MFR) and single-fidelity regression trained on HF data only (SFR).}
    \label{fig:C}
\end{figure}
Here, $51$ and $15$ equally spaced points in $x \in [0,1]$ are used as low- and high-fidelity observed locations with learning rate of $0.01$, and the second network is trained with $\lambda=0.0001$. The results of Fig.~\ref{fig:C} indicate that both multi-fidelity models (GH-MFR and MFR) are able to accurately capture non-linear relationships and achieve results that are indistinguishable from one another. Furthermore, these models are more precise than single-fidelity model (SFR) trained only on high-fidelity data.

\subsection{Phase-shifted oscillations}
\begin{figure}
    \centering
    \includegraphics[width=0.5\textwidth]{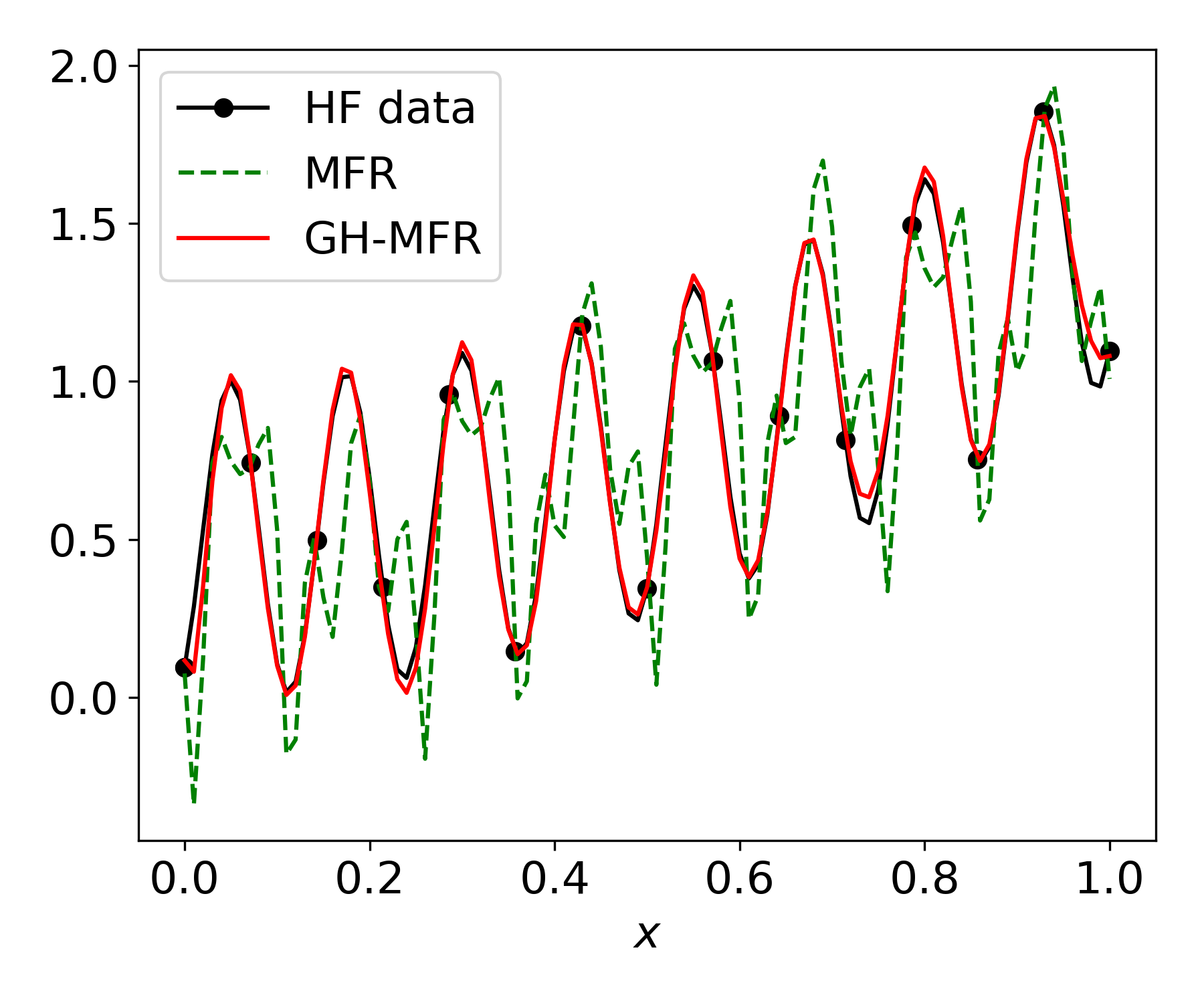}
    \caption{Results of gradient-enhanced multi-fidelity regression (GH-MFR) for case D with $51$ low-fidelity (LF) and $15$ high-fidelity (HF) observations. Performance is compared to multi-fidelity regression without gradient information (MFR) and single-fidelity regression trained on HF data only (SFR).}
    \label{fig:D2}
\end{figure}

The final benchmark case demonstrates the importance of using gradient information in cases where more complex relations exist between low- and high-fidelity data. In this case, the data are obtained from,
\begin{align*}
    y_{LF} &= \sin (8 \pi x),  \\
    y_{HF} &= x^2 + \sin^2 (8 \pi x + \frac{\pi}{10}),
\end{align*}
\begin{figure}
    \centering
    \includegraphics[width=0.5\textwidth]{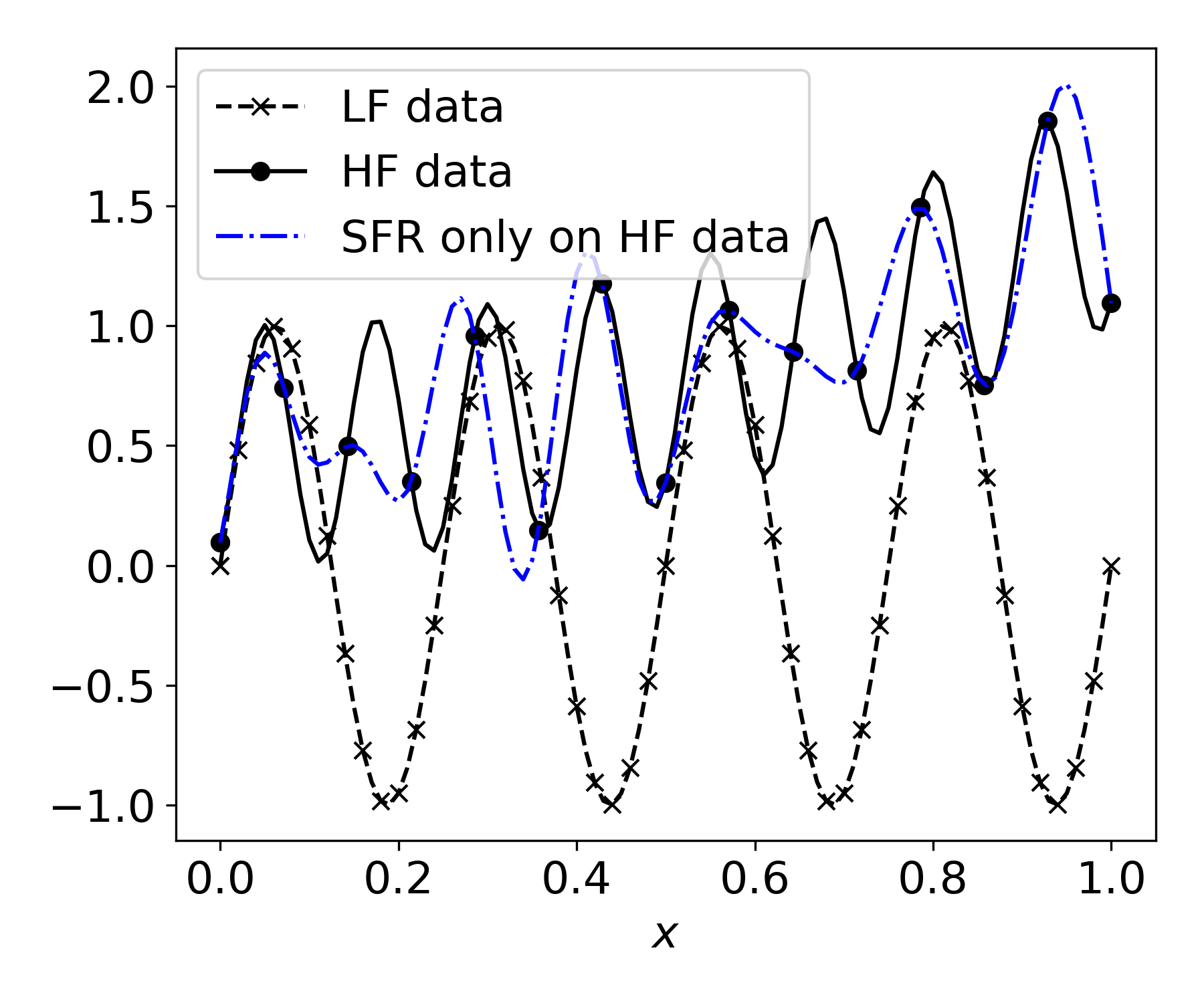}
    \caption{Single-fidelity regression (SFR) results for Case D with observations at $51$ low-fidelity (LF) and $15$ high-fidelity (HF) locations.}
    \label{fig:D1}
\end{figure}
with $51$ and $15$ equally spaced observed locations for $\bm{x}_{LF}$ and $\bm{x}_{HF}$ shown in Fig.~\ref{fig:D1} along with the single-fidelity model (SFR) through high-fidelity points. The same learning and regularization rates employed in the previous case were also utilized in this experiment. Figure~\ref{fig:D2} presents the results obtained by multi-fidelity models. It is clearly observed that the vanilla multi-fidelity model without gradient information (MFR) fails to represent the high-fidelity signal, while the gradient enhanced model (GH-MFR) is successful in providing an accurate prediction of the high-fidelity data.

\section{Turbulent flow reconstruction} \label{Sec:Turb}
Despite unprecedented progress in computing power as well as experimental setups, obtaining high-fidelity fluid flow data remains a difficult and demanding task. This challenge becomes even more prominent in industrial relevant applications, where flow becomes turbulent. A wide range of length and time scales in turbulent flows makes it prohibitively expensive to conduct fully-resolved direct numerical simulation (DNS) or experimental measurement. Nevertheless, the abundance of low-resolution flow data, through coarse numerical simulation or limited sensor measurements, opens the possibility of building data-driven models to generate fast and reliable high-resolution data. 

The application of data-driven techniques in fluid dynamics application is a rapidly growing field \cite{brunton2020machine, vinuesa2022emerging, vinuesa2022enhancing}. Data-driven methods, have been used to speed-up numerical simulations, enahnce existing turbulence models and build non-linear reduced-order models, to name a few \cite{vinuesa2022enhancing}. Additionally, a number of data-driven models have been proposed in the field of flow reconstruction. Traditional models are based on proper-orthogonal decomposition (POD) or compressed sensing techniques \cite{erichson2020shallow, donoho2006compressed}, which are mainly linear and not suitable for accurate reconstruction of turbulent chaotic field \cite{vinuesa2022emerging}. Another family of models based on neural networks and deep learning techniques, however, are capable of capturing complex non-linear interactions. In particular, various super-resolution models based on convolutional neural networks (CNNs) or generative adversarial networks (GANs) have shown promising results for reconstructing high-resolution data from low-resolution data \cite{deng2019super, fukami2019super, liu2020deep, liu2020deep, gao2021super, yousif2021high, matsuo2021supervised,guemes2021coarse, zhou2022robust, yousif2022super, yu2022three}. Nevertheless, the main limitation of these models is that they require a considerable amount of high-fidelity/high-resolution data during the training phase, which may be difficult to acquire in many practical realistic scenarios.   
\begin{figure*}[]
    \centering
    \includegraphics[width=1.0\textwidth]{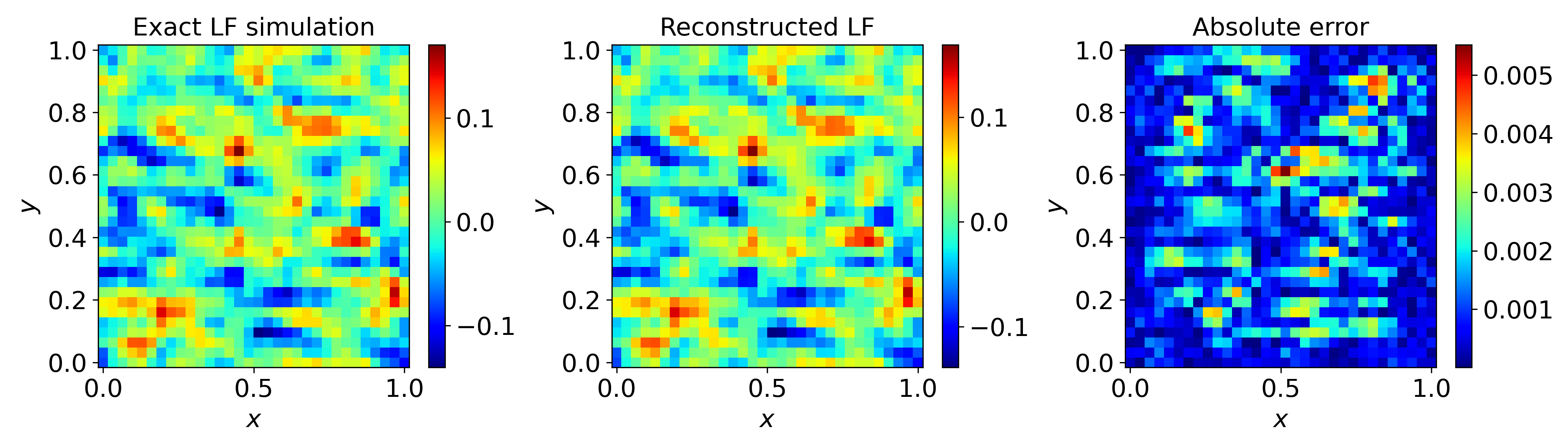}
    \caption{The contours of $u$-velocity of isotropic turbulence for low-fidelity simulation (left) and its reconstruction with $\mathcal{NN}_{LF}$ (middle).}
    \label{fig:Decay_LF}
\end{figure*}
\begin{figure*}[]
    \centering
    \includegraphics[width=1.0\textwidth]{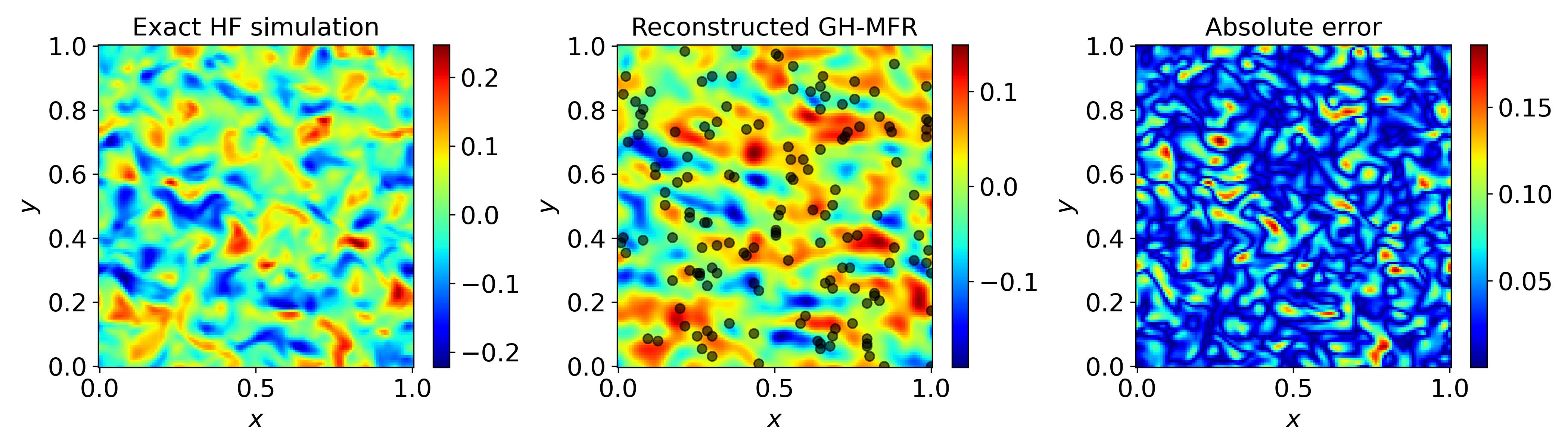}
    \includegraphics[width=1.0\textwidth]{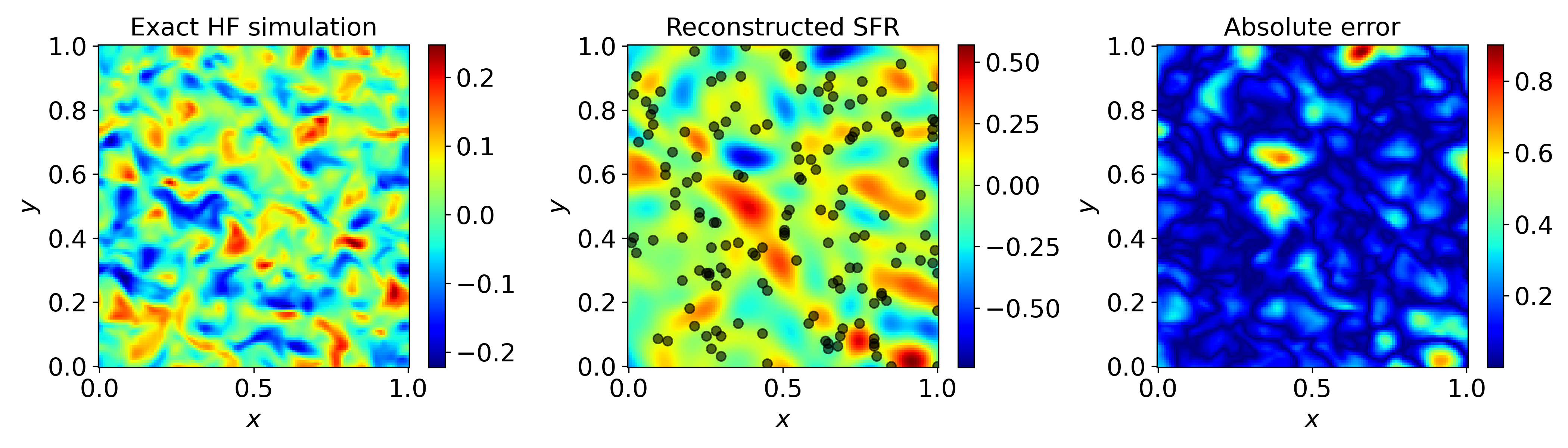}
    \caption{The contours of $u$-velocity of isotropic turbulence for high-fidelity simulation (left) and its reconstruction with the gradient enhanced multi-fidelity model (top-middle) and with the single-fidelity model (bottom-middle). Black circles represent the high-fidelity sample points.}
    \label{fig:Decay_MF}
\end{figure*}
Multi-fidelity models can circumvent this limitation as they rely on few high-fidelity/high-resolution observation points \cite{conti2023multi, deng2020multifidelity, pawar2022multi, zhang2022fusion, mondal2022multi}. 

Building on the results of the previous section, this section demonstrates the effectiveness of the proposed gradient-enhanced multi-fidelity model for a practical and challenging case of turbulent flow reconstruction.
\begin{figure*}[]
    \centering
    \includegraphics[width=1.0\textwidth]{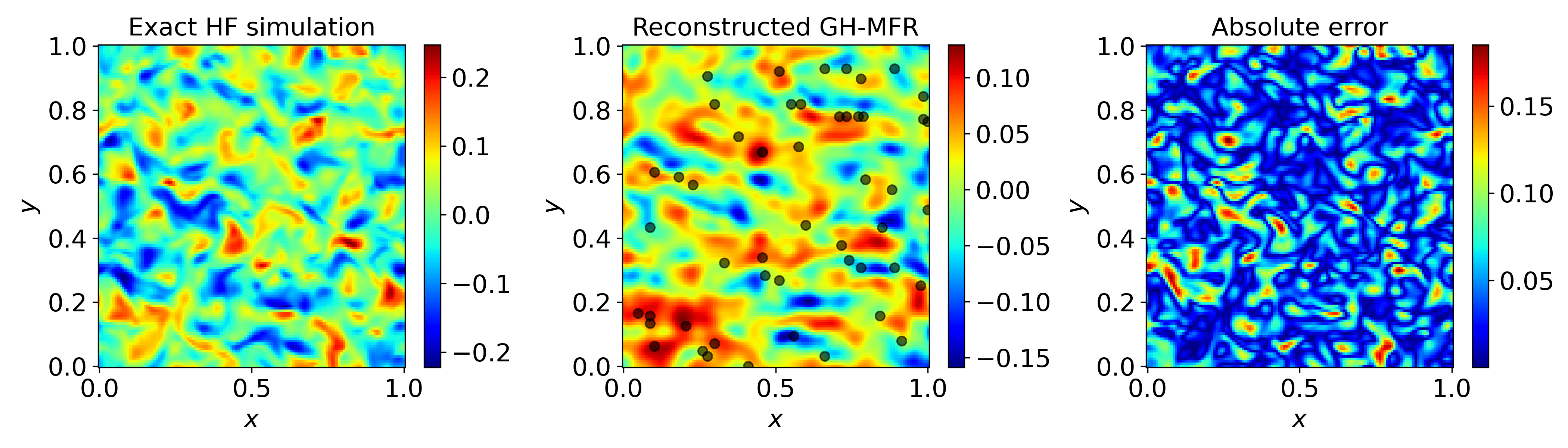}
    \caption{The contours of $u$-velocity of isotropic turbulence for high-fidelity simulation (left) and its reconstruction with the gradient enhanced multi-fidelity model (middle) in extremely low-data regime. Black circles represent the high-fidelity sample points.}
    \label{fig:Decay_LowData}
\end{figure*}

The decaying of a compressible homogeneous isotropic turbulence in a periodic box $(x,y,z) \in [0,1]$ is studied, with the turbulent Mach number $\text{Ma}_t=0.3$ and the Reynolds number based on the the Taylor microscale of $\text{Re}_\lambda=72$. Further details on the
numerical setup can be found in \cite{samtaney2001direct}.This case is challenging due to the presence of compressibility and turbulent structures in the flow field \cite{samtaney2001direct,johnsen2010assessment}. The reconstruction of the instantaneous $x-$component of the velocity field $\bm{u}=(u, v, w)$ on a $2D$ slice at $z=0.5$ is investigated. 

Training data are generated by solving a $3D$ compressible lattice Boltzmann (LB) model \cite{saadat2021extended, saadat2019lattice} on uniformly spaced grid of size $32^3$ and $128^3$. The LB model recovers the following Navier-Stokes-Fourier (NSF) equations for the density $\rho$, momentum $\rho \bm{u}$ and energy $\rho E$,
\begin{align}
    &\partial_t\rho + \nabla\cdot(\rho\bm{u}) = 0,\label{eq:continuity} 
    \\
    &\partial_t (\rho \bm{u})
    + \nabla \cdot (\rho \bm{u}\otimes\bm{u}) + \nabla\cdot \bm{\pi} = 0,
	\label{eq:flow}
	\\
	&\partial_t (\rho E) + \nabla\cdot(\rho E \bm{u})
    + \nabla \cdot \bm{q} + \nabla\cdot (\bm{\pi}\cdot\bm{u}) = 0. \label{eq:energyeq}
\end{align}
Here, $\bm{\pi}$ is the pressure tensor,
\begin{align}
    \bm{\pi} = P\bm{I} - \mu \left(\bm{S} - \frac{2}{3}(\nabla\cdot\bm{u})\bm{I} \right) - \varsigma(\nabla\cdot\bm{u})\bm{I}, 
\end{align}
with $P=\rho RT$ being the pressure of ideal gas, $\mu$ the dynamic viscosity, $\varsigma$ the bulk viscosity,
and the strain rate tensor is
\begin{align}
    \bm{S} = \nabla \bm{u} + \nabla \bm{u}^\dagger. \label{eq:strain}
\end{align}
All $32^2$ points in the $2D$ slice are assumed as low-fidelity observation points, while $150$ points are randomly sampled from the high-resolution simulation (i.e., from $128^2$ points) as sparse high-fidelity observations (see black circles in Fig.~\ref{fig:Decay_MF} middle). Here, the network architecture of the multi-fidelity model is with $3$ hidden layers, $200$ neurons in each layer and $\sin$ activation function. The regularization rate of the second network is set to $\lambda = 0.0008$. 

Figure~\ref{fig:Decay_LF} shows the instantaneous $u$-velocity field of low-fidelity simulation along with its reconstructed field using the low-fidelity network (i.e., $\mathcal{NN}_{LF}$). Now, by taking advantage of the reconstructed low-fidelity information at sparse high-fidelity points, the results of the gradient enhanced multi-fidelity model (GH-MFR) is computed with the second network (i.e., $\mathcal{NN}_{HF}$) and illustrated in Fig.~\ref{fig:Decay_MF}, in comparison with the exact high-fidelity data from numerical simulation. The reconstructed field indicates that the present model is capable of recovering the exact field with reasonable detail and accuracy where the relative error measured in the $L_2$ norm is $0.0027$. On the other hand, the single-fidelity regression through only high-fidelity data points shown in Fig.~\ref{fig:Decay_MF} bottom, results in a poor reconstruction that over-smooths the flow field and is unable to capture any small scale structures. The $L_2$ norm error in this case is $0.028$. 

To further assess the capability of the proposed multi-fidelity model, we predict the reconstructed field in an extremely low-data regime where only $50$ random points are used as high-fidelity observations. The resulting flow field shown in Fig.~\ref{fig:Decay_LowData} with $L_2$ error of $0.0033$ demonstrates the superior performance of multi-fidelity model compared to single-fidelity models even in extreme cases with very sparse high-fidelity observations.  

\section{Conclusions} \label{Sec:conclusion}
Multi-fidelity modeling seeks to leverage data-driven techniques to combine abundant, inexpensive low-fidelity data with scarce, costly high-fidelity data, taking advantage of the correlation between them to develop a surrogate model that outperforms the model that relies solely on a few high-fidelity data points.

In this paper, A multi-fidelity regression model based on neural networks \cite{} was proposed. Inspired by the "2-step" neural network multi-fidelity model \cite{guo2022multi, meng2020composite}, the present model utilizes two separate networks for learning from low- and high-fidelity data and further enhances the model by feeding the gradients information of low-fidelity data into the second network. The gradients are computed using the automatic differentiation with minimal computational effort. The accuracy of the proposed gradient-enhanced multi-fidelity framework was demonstrated through a series of synthetic benchmark tests, and it was found that the proposed model yields better results compared to traditional "2-step" multi-fidelity model that lacks gradient information. 

In the next step, the effectiveness of the proposed multi-fidelity model  was examined in a challenging case of reconstructing the instantaneous velocity field of a compressible homogeneous isotropic turbulence given low-resolution/low-fidelity data and a small amount of high-resolution/high-fidelity data. The results showed that the model is able to capture the fine scale structures of the turbulent field.

The promising results of the proposed multi-fidelity model, make it a viable approach for producing high-resolution data in scenarios where sparse experimental measurements and a reliable low-fidelity model are available. This could be a beneficial tool for various realistic practical applications and will be investigated in future studies.

\begin{acknowledgments}
The author would like to thank Nikolaos Kallikounis for helpful discussions.
\end{acknowledgments}


\bibliography{Bib}

\begin{thebibliography}{41}%
\makeatletter
\providecommand \@ifxundefined [1]{%
 \@ifx{#1\undefined}
}%
\providecommand \@ifnum [1]{%
 \ifnum #1\expandafter \@firstoftwo
 \else \expandafter \@secondoftwo
 \fi
}%
\providecommand \@ifx [1]{%
 \ifx #1\expandafter \@firstoftwo
 \else \expandafter \@secondoftwo
 \fi
}%
\providecommand \natexlab [1]{#1}%
\providecommand \enquote  [1]{``#1''}%
\providecommand \bibnamefont  [1]{#1}%
\providecommand \bibfnamefont [1]{#1}%
\providecommand \citenamefont [1]{#1}%
\providecommand \href@noop [0]{\@secondoftwo}%
\providecommand \href [0]{\begingroup \@sanitize@url \@href}%
\providecommand \@href[1]{\@@startlink{#1}\@@href}%
\providecommand \@@href[1]{\endgroup#1\@@endlink}%
\providecommand \@sanitize@url [0]{\catcode `\\12\catcode `\$12\catcode
  `\&12\catcode `\#12\catcode `\^12\catcode `\_12\catcode `\%12\relax}%
\providecommand \@@startlink[1]{}%
\providecommand \@@endlink[0]{}%
\providecommand \url  [0]{\begingroup\@sanitize@url \@url }%
\providecommand \@url [1]{\endgroup\@href {#1}{\urlprefix }}%
\providecommand \urlprefix  [0]{URL }%
\providecommand \Eprint [0]{\href }%
\providecommand \doibase [0]{http://dx.doi.org/}%
\providecommand \selectlanguage [0]{\@gobble}%
\providecommand \bibinfo  [0]{\@secondoftwo}%
\providecommand \bibfield  [0]{\@secondoftwo}%
\providecommand \translation [1]{[#1]}%
\providecommand \BibitemOpen [0]{}%
\providecommand \bibitemStop [0]{}%
\providecommand \bibitemNoStop [0]{.\EOS\space}%
\providecommand \EOS [0]{\spacefactor3000\relax}%
\providecommand \BibitemShut  [1]{\csname bibitem#1\endcsname}%
\let\auto@bib@innerbib\@empty
\bibitem [{\citenamefont {Meng}\ and\ \citenamefont
  {Karniadakis}(2020)}]{meng2020composite}%
  \BibitemOpen
  \bibfield  {author} {\bibinfo {author} {\bibfnamefont {X.}~\bibnamefont
  {Meng}}\ and\ \bibinfo {author} {\bibfnamefont {G.~E.}\ \bibnamefont
  {Karniadakis}},\ }\bibfield  {title} {\enquote {\bibinfo {title} {A composite
  neural network that learns from multi-fidelity data: Application to function
  approximation and inverse pde problems},}\ }\href@noop {} {\bibfield
  {journal} {\bibinfo  {journal} {Journal of Computational Physics}\ }\textbf
  {\bibinfo {volume} {401}},\ \bibinfo {pages} {109020} (\bibinfo {year}
  {2020})}\BibitemShut {NoStop}%
\bibitem [{\citenamefont {Perdikaris}\ \emph {et~al.}(2017)\citenamefont
  {Perdikaris}, \citenamefont {Raissi}, \citenamefont {Damianou}, \citenamefont
  {Lawrence},\ and\ \citenamefont {Karniadakis}}]{perdikaris2017nonlinear}%
  \BibitemOpen
  \bibfield  {author} {\bibinfo {author} {\bibfnamefont {P.}~\bibnamefont
  {Perdikaris}}, \bibinfo {author} {\bibfnamefont {M.}~\bibnamefont {Raissi}},
  \bibinfo {author} {\bibfnamefont {A.}~\bibnamefont {Damianou}}, \bibinfo
  {author} {\bibfnamefont {N.~D.}\ \bibnamefont {Lawrence}}, \ and\ \bibinfo
  {author} {\bibfnamefont {G.~E.}\ \bibnamefont {Karniadakis}},\ }\bibfield
  {title} {\enquote {\bibinfo {title} {Nonlinear information fusion algorithms
  for data-efficient multi-fidelity modelling},}\ }\href@noop {} {\bibfield
  {journal} {\bibinfo  {journal} {Proceedings of the Royal Society A:
  Mathematical, Physical and Engineering Sciences}\ }\textbf {\bibinfo {volume}
  {473}},\ \bibinfo {pages} {20160751} (\bibinfo {year} {2017})}\BibitemShut
  {NoStop}%
\bibitem [{\citenamefont {Parussini}\ \emph {et~al.}(2017)\citenamefont
  {Parussini}, \citenamefont {Venturi}, \citenamefont {Perdikaris},\ and\
  \citenamefont {Karniadakis}}]{parussini2017multi}%
  \BibitemOpen
  \bibfield  {author} {\bibinfo {author} {\bibfnamefont {L.}~\bibnamefont
  {Parussini}}, \bibinfo {author} {\bibfnamefont {D.}~\bibnamefont {Venturi}},
  \bibinfo {author} {\bibfnamefont {P.}~\bibnamefont {Perdikaris}}, \ and\
  \bibinfo {author} {\bibfnamefont {G.~E.}\ \bibnamefont {Karniadakis}},\
  }\bibfield  {title} {\enquote {\bibinfo {title} {Multi-fidelity gaussian
  process regression for prediction of random fields},}\ }\href@noop {}
  {\bibfield  {journal} {\bibinfo  {journal} {Journal of Computational
  Physics}\ }\textbf {\bibinfo {volume} {336}},\ \bibinfo {pages} {36--50}
  (\bibinfo {year} {2017})}\BibitemShut {NoStop}%
\bibitem [{\citenamefont {Lee}\ \emph {et~al.}(2019)\citenamefont {Lee},
  \citenamefont {Dietrich}, \citenamefont {Karniadakis},\ and\ \citenamefont
  {Kevrekidis}}]{lee2019linking}%
  \BibitemOpen
  \bibfield  {author} {\bibinfo {author} {\bibfnamefont {S.}~\bibnamefont
  {Lee}}, \bibinfo {author} {\bibfnamefont {F.}~\bibnamefont {Dietrich}},
  \bibinfo {author} {\bibfnamefont {G.~E.}\ \bibnamefont {Karniadakis}}, \ and\
  \bibinfo {author} {\bibfnamefont {I.~G.}\ \bibnamefont {Kevrekidis}},\
  }\bibfield  {title} {\enquote {\bibinfo {title} {Linking gaussian process
  regression with data-driven manifold embeddings for nonlinear data fusion},}\
  }\href@noop {} {\bibfield  {journal} {\bibinfo  {journal} {Interface focus}\
  }\textbf {\bibinfo {volume} {9}},\ \bibinfo {pages} {20180083} (\bibinfo
  {year} {2019})}\BibitemShut {NoStop}%
\bibitem [{\citenamefont {Deng}, \citenamefont {Lin},\ and\ \citenamefont
  {Yang}(2020)}]{deng2020multifidelity}%
  \BibitemOpen
  \bibfield  {author} {\bibinfo {author} {\bibfnamefont {Y.}~\bibnamefont
  {Deng}}, \bibinfo {author} {\bibfnamefont {G.}~\bibnamefont {Lin}}, \ and\
  \bibinfo {author} {\bibfnamefont {X.}~\bibnamefont {Yang}},\ }\bibfield
  {title} {\enquote {\bibinfo {title} {Multifidelity data fusion via
  gradient-enhanced gaussian process regression},}\ }\href@noop {} {\bibfield
  {journal} {\bibinfo  {journal} {arXiv preprint arXiv:2008.01066}\ } (\bibinfo
  {year} {2020})}\BibitemShut {NoStop}%
\bibitem [{\citenamefont {Stanek}, \citenamefont {Bopardikar},\ and\
  \citenamefont {Murillo}(2021)}]{stanek2021multifidelity}%
  \BibitemOpen
  \bibfield  {author} {\bibinfo {author} {\bibfnamefont {L.~J.}\ \bibnamefont
  {Stanek}}, \bibinfo {author} {\bibfnamefont {S.~D.}\ \bibnamefont
  {Bopardikar}}, \ and\ \bibinfo {author} {\bibfnamefont {M.~S.}\ \bibnamefont
  {Murillo}},\ }\bibfield  {title} {\enquote {\bibinfo {title} {Multifidelity
  regression of sparse plasma transport data available in disparate physical
  regimes},}\ }\href@noop {} {\bibfield  {journal} {\bibinfo  {journal}
  {Physical Review E}\ }\textbf {\bibinfo {volume} {104}},\ \bibinfo {pages}
  {065303} (\bibinfo {year} {2021})}\BibitemShut {NoStop}%
\bibitem [{\citenamefont {Raissi}, \citenamefont {Perdikaris},\ and\
  \citenamefont {Karniadakis}(2019)}]{raissi2019physics}%
  \BibitemOpen
  \bibfield  {author} {\bibinfo {author} {\bibfnamefont {M.}~\bibnamefont
  {Raissi}}, \bibinfo {author} {\bibfnamefont {P.}~\bibnamefont {Perdikaris}},
  \ and\ \bibinfo {author} {\bibfnamefont {G.~E.}\ \bibnamefont
  {Karniadakis}},\ }\bibfield  {title} {\enquote {\bibinfo {title}
  {Physics-informed neural networks: A deep learning framework for solving
  forward and inverse problems involving nonlinear partial differential
  equations},}\ }\href@noop {} {\bibfield  {journal} {\bibinfo  {journal}
  {Journal of Computational physics}\ }\textbf {\bibinfo {volume} {378}},\
  \bibinfo {pages} {686--707} (\bibinfo {year} {2019})}\BibitemShut {NoStop}%
\bibitem [{\citenamefont {Eivazi}\ \emph {et~al.}(2022)\citenamefont {Eivazi},
  \citenamefont {Tahani}, \citenamefont {Schlatter},\ and\ \citenamefont
  {Vinuesa}}]{eivazi2022physics}%
  \BibitemOpen
  \bibfield  {author} {\bibinfo {author} {\bibfnamefont {H.}~\bibnamefont
  {Eivazi}}, \bibinfo {author} {\bibfnamefont {M.}~\bibnamefont {Tahani}},
  \bibinfo {author} {\bibfnamefont {P.}~\bibnamefont {Schlatter}}, \ and\
  \bibinfo {author} {\bibfnamefont {R.}~\bibnamefont {Vinuesa}},\ }\bibfield
  {title} {\enquote {\bibinfo {title} {Physics-informed neural networks for
  solving reynolds-averaged navier--stokes equations},}\ }\href@noop {}
  {\bibfield  {journal} {\bibinfo  {journal} {Physics of Fluids}\ }\textbf
  {\bibinfo {volume} {34}},\ \bibinfo {pages} {075117} (\bibinfo {year}
  {2022})}\BibitemShut {NoStop}%
\bibitem [{\citenamefont {Raissi}\ and\ \citenamefont
  {Karniadakis}(2018)}]{raissi2018hidden}%
  \BibitemOpen
  \bibfield  {author} {\bibinfo {author} {\bibfnamefont {M.}~\bibnamefont
  {Raissi}}\ and\ \bibinfo {author} {\bibfnamefont {G.~E.}\ \bibnamefont
  {Karniadakis}},\ }\bibfield  {title} {\enquote {\bibinfo {title} {Hidden
  physics models: Machine learning of nonlinear partial differential
  equations},}\ }\href@noop {} {\bibfield  {journal} {\bibinfo  {journal}
  {Journal of Computational Physics}\ }\textbf {\bibinfo {volume} {357}},\
  \bibinfo {pages} {125--141} (\bibinfo {year} {2018})}\BibitemShut {NoStop}%
\bibitem [{\citenamefont {Chen}, \citenamefont {Liu},\ and\ \citenamefont
  {Sun}(2021)}]{chen2021physics}%
  \BibitemOpen
  \bibfield  {author} {\bibinfo {author} {\bibfnamefont {Z.}~\bibnamefont
  {Chen}}, \bibinfo {author} {\bibfnamefont {Y.}~\bibnamefont {Liu}}, \ and\
  \bibinfo {author} {\bibfnamefont {H.}~\bibnamefont {Sun}},\ }\bibfield
  {title} {\enquote {\bibinfo {title} {Physics-informed learning of governing
  equations from scarce data},}\ }\href@noop {} {\bibfield  {journal} {\bibinfo
   {journal} {Nature communications}\ }\textbf {\bibinfo {volume} {12}},\
  \bibinfo {pages} {1--13} (\bibinfo {year} {2021})}\BibitemShut {NoStop}%
\bibitem [{\citenamefont {Fukami}\ \emph {et~al.}(2021)\citenamefont {Fukami},
  \citenamefont {Hasegawa}, \citenamefont {Nakamura}, \citenamefont
  {Morimoto},\ and\ \citenamefont {Fukagata}}]{fukami2021model}%
  \BibitemOpen
  \bibfield  {author} {\bibinfo {author} {\bibfnamefont {K.}~\bibnamefont
  {Fukami}}, \bibinfo {author} {\bibfnamefont {K.}~\bibnamefont {Hasegawa}},
  \bibinfo {author} {\bibfnamefont {T.}~\bibnamefont {Nakamura}}, \bibinfo
  {author} {\bibfnamefont {M.}~\bibnamefont {Morimoto}}, \ and\ \bibinfo
  {author} {\bibfnamefont {K.}~\bibnamefont {Fukagata}},\ }\bibfield  {title}
  {\enquote {\bibinfo {title} {Model order reduction with neural networks:
  Application to laminar and turbulent flows},}\ }\href@noop {} {\bibfield
  {journal} {\bibinfo  {journal} {SN Computer Science}\ }\textbf {\bibinfo
  {volume} {2}},\ \bibinfo {pages} {1--16} (\bibinfo {year}
  {2021})}\BibitemShut {NoStop}%
\bibitem [{\citenamefont {Guo}\ \emph {et~al.}(2022)\citenamefont {Guo},
  \citenamefont {Manzoni}, \citenamefont {Amendt}, \citenamefont {Conti},\ and\
  \citenamefont {Hesthaven}}]{guo2022multi}%
  \BibitemOpen
  \bibfield  {author} {\bibinfo {author} {\bibfnamefont {M.}~\bibnamefont
  {Guo}}, \bibinfo {author} {\bibfnamefont {A.}~\bibnamefont {Manzoni}},
  \bibinfo {author} {\bibfnamefont {M.}~\bibnamefont {Amendt}}, \bibinfo
  {author} {\bibfnamefont {P.}~\bibnamefont {Conti}}, \ and\ \bibinfo {author}
  {\bibfnamefont {J.~S.}\ \bibnamefont {Hesthaven}},\ }\bibfield  {title}
  {\enquote {\bibinfo {title} {Multi-fidelity regression using artificial
  neural networks: efficient approximation of parameter-dependent output
  quantities},}\ }\href@noop {} {\bibfield  {journal} {\bibinfo  {journal}
  {Computer methods in applied mechanics and engineering}\ }\textbf {\bibinfo
  {volume} {389}},\ \bibinfo {pages} {114378} (\bibinfo {year}
  {2022})}\BibitemShut {NoStop}%
\bibitem [{\citenamefont {Conti}\ \emph {et~al.}(2023)\citenamefont {Conti},
  \citenamefont {Guo}, \citenamefont {Manzoni},\ and\ \citenamefont
  {Hesthaven}}]{conti2023multi}%
  \BibitemOpen
  \bibfield  {author} {\bibinfo {author} {\bibfnamefont {P.}~\bibnamefont
  {Conti}}, \bibinfo {author} {\bibfnamefont {M.}~\bibnamefont {Guo}}, \bibinfo
  {author} {\bibfnamefont {A.}~\bibnamefont {Manzoni}}, \ and\ \bibinfo
  {author} {\bibfnamefont {J.~S.}\ \bibnamefont {Hesthaven}},\ }\bibfield
  {title} {\enquote {\bibinfo {title} {Multi-fidelity surrogate modeling using
  long short-term memory networks},}\ }\href@noop {} {\bibfield  {journal}
  {\bibinfo  {journal} {Computer methods in applied mechanics and engineering}\
  }\textbf {\bibinfo {volume} {404}},\ \bibinfo {pages} {115811} (\bibinfo
  {year} {2023})}\BibitemShut {NoStop}%
\bibitem [{\citenamefont {Partin}\ \emph {et~al.}(2023)\citenamefont {Partin},
  \citenamefont {Geraci}, \citenamefont {Rushdi}, \citenamefont {Eldred},\ and\
  \citenamefont {Schiavazzi}}]{partin2023multifidelity}%
  \BibitemOpen
  \bibfield  {author} {\bibinfo {author} {\bibfnamefont {L.}~\bibnamefont
  {Partin}}, \bibinfo {author} {\bibfnamefont {G.}~\bibnamefont {Geraci}},
  \bibinfo {author} {\bibfnamefont {A.~A.}\ \bibnamefont {Rushdi}}, \bibinfo
  {author} {\bibfnamefont {M.~S.}\ \bibnamefont {Eldred}}, \ and\ \bibinfo
  {author} {\bibfnamefont {D.~E.}\ \bibnamefont {Schiavazzi}},\ }\bibfield
  {title} {\enquote {\bibinfo {title} {Multifidelity data fusion in
  convolutional encoder/decoder networks},}\ }\href@noop {} {\bibfield
  {journal} {\bibinfo  {journal} {Journal of Computational Physics}\ }\textbf
  {\bibinfo {volume} {472}},\ \bibinfo {pages} {111666} (\bibinfo {year}
  {2023})}\BibitemShut {NoStop}%
\bibitem [{\citenamefont {Pawar}\ \emph {et~al.}(2022)\citenamefont {Pawar},
  \citenamefont {San}, \citenamefont {Vedula}, \citenamefont {Rasheed},\ and\
  \citenamefont {Kvamsdal}}]{pawar2022multi}%
  \BibitemOpen
  \bibfield  {author} {\bibinfo {author} {\bibfnamefont {S.}~\bibnamefont
  {Pawar}}, \bibinfo {author} {\bibfnamefont {O.}~\bibnamefont {San}}, \bibinfo
  {author} {\bibfnamefont {P.}~\bibnamefont {Vedula}}, \bibinfo {author}
  {\bibfnamefont {A.}~\bibnamefont {Rasheed}}, \ and\ \bibinfo {author}
  {\bibfnamefont {T.}~\bibnamefont {Kvamsdal}},\ }\bibfield  {title} {\enquote
  {\bibinfo {title} {Multi-fidelity information fusion with concatenated neural
  networks},}\ }\href@noop {} {\bibfield  {journal} {\bibinfo  {journal}
  {Scientific Reports}\ }\textbf {\bibinfo {volume} {12}},\ \bibinfo {pages}
  {1--13} (\bibinfo {year} {2022})}\BibitemShut {NoStop}%
\bibitem [{\citenamefont {Meng}, \citenamefont {Babaee},\ and\ \citenamefont
  {Karniadakis}(2021)}]{meng2021multi}%
  \BibitemOpen
  \bibfield  {author} {\bibinfo {author} {\bibfnamefont {X.}~\bibnamefont
  {Meng}}, \bibinfo {author} {\bibfnamefont {H.}~\bibnamefont {Babaee}}, \ and\
  \bibinfo {author} {\bibfnamefont {G.~E.}\ \bibnamefont {Karniadakis}},\
  }\bibfield  {title} {\enquote {\bibinfo {title} {Multi-fidelity bayesian
  neural networks: Algorithms and applications},}\ }\href@noop {} {\bibfield
  {journal} {\bibinfo  {journal} {Journal of Computational Physics}\ }\textbf
  {\bibinfo {volume} {438}},\ \bibinfo {pages} {110361} (\bibinfo {year}
  {2021})}\BibitemShut {NoStop}%
\bibitem [{\citenamefont {Sopena}, \citenamefont {Romero},\ and\ \citenamefont
  {Alquezar}(1999)}]{sopena1999neural}%
  \BibitemOpen
  \bibfield  {author} {\bibinfo {author} {\bibfnamefont {J.~M.}\ \bibnamefont
  {Sopena}}, \bibinfo {author} {\bibfnamefont {E.}~\bibnamefont {Romero}}, \
  and\ \bibinfo {author} {\bibfnamefont {R.}~\bibnamefont {Alquezar}},\
  }\bibfield  {title} {\enquote {\bibinfo {title} {Neural networks with
  periodic and monotonic activation functions: a comparative study in
  classification problems},}\ }\href@noop {} {\  (\bibinfo {year}
  {1999})}\BibitemShut {NoStop}%
\bibitem [{\citenamefont {Parascandolo}, \citenamefont {Huttunen},\ and\
  \citenamefont {Virtanen}(2016)}]{parascandolo2016taming}%
  \BibitemOpen
  \bibfield  {author} {\bibinfo {author} {\bibfnamefont {G.}~\bibnamefont
  {Parascandolo}}, \bibinfo {author} {\bibfnamefont {H.}~\bibnamefont
  {Huttunen}}, \ and\ \bibinfo {author} {\bibfnamefont {T.}~\bibnamefont
  {Virtanen}},\ }\bibfield  {title} {\enquote {\bibinfo {title} {Taming the
  waves: sine as activation function in deep neural networks},}\ }\href@noop {}
  {\  (\bibinfo {year} {2016})}\BibitemShut {NoStop}%
\bibitem [{\citenamefont {Sitzmann}\ \emph {et~al.}(2020)\citenamefont
  {Sitzmann}, \citenamefont {Martel}, \citenamefont {Bergman}, \citenamefont
  {Lindell},\ and\ \citenamefont {Wetzstein}}]{sitzmann2020implicit}%
  \BibitemOpen
  \bibfield  {author} {\bibinfo {author} {\bibfnamefont {V.}~\bibnamefont
  {Sitzmann}}, \bibinfo {author} {\bibfnamefont {J.}~\bibnamefont {Martel}},
  \bibinfo {author} {\bibfnamefont {A.}~\bibnamefont {Bergman}}, \bibinfo
  {author} {\bibfnamefont {D.}~\bibnamefont {Lindell}}, \ and\ \bibinfo
  {author} {\bibfnamefont {G.}~\bibnamefont {Wetzstein}},\ }\bibfield  {title}
  {\enquote {\bibinfo {title} {Implicit neural representations with periodic
  activation functions},}\ }\href@noop {} {\bibfield  {journal} {\bibinfo
  {journal} {Advances in Neural Information Processing Systems}\ }\textbf
  {\bibinfo {volume} {33}},\ \bibinfo {pages} {7462--7473} (\bibinfo {year}
  {2020})}\BibitemShut {NoStop}%
\bibitem [{\citenamefont {Wong}\ \emph {et~al.}(2022)\citenamefont {Wong},
  \citenamefont {Ooi}, \citenamefont {Gupta},\ and\ \citenamefont
  {Ong}}]{wong2022learning}%
  \BibitemOpen
  \bibfield  {author} {\bibinfo {author} {\bibfnamefont {J.~C.}\ \bibnamefont
  {Wong}}, \bibinfo {author} {\bibfnamefont {C.}~\bibnamefont {Ooi}}, \bibinfo
  {author} {\bibfnamefont {A.}~\bibnamefont {Gupta}}, \ and\ \bibinfo {author}
  {\bibfnamefont {Y.-S.}\ \bibnamefont {Ong}},\ }\bibfield  {title} {\enquote
  {\bibinfo {title} {Learning in sinusoidal spaces with physics-informed neural
  networks},}\ }\href@noop {} {\bibfield  {journal} {\bibinfo  {journal} {IEEE
  Transactions on Artificial Intelligence}\ } (\bibinfo {year}
  {2022})}\BibitemShut {NoStop}%
\bibitem [{\citenamefont {Brunton}, \citenamefont {Noack},\ and\ \citenamefont
  {Koumoutsakos}(2020)}]{brunton2020machine}%
  \BibitemOpen
  \bibfield  {author} {\bibinfo {author} {\bibfnamefont {S.~L.}\ \bibnamefont
  {Brunton}}, \bibinfo {author} {\bibfnamefont {B.~R.}\ \bibnamefont {Noack}},
  \ and\ \bibinfo {author} {\bibfnamefont {P.}~\bibnamefont {Koumoutsakos}},\
  }\bibfield  {title} {\enquote {\bibinfo {title} {Machine learning for fluid
  mechanics},}\ }\href@noop {} {\bibfield  {journal} {\bibinfo  {journal}
  {Annual review of fluid mechanics}\ }\textbf {\bibinfo {volume} {52}},\
  \bibinfo {pages} {477--508} (\bibinfo {year} {2020})}\BibitemShut {NoStop}%
\bibitem [{\citenamefont {Vinuesa}\ and\ \citenamefont
  {Brunton}(2022{\natexlab{a}})}]{vinuesa2022emerging}%
  \BibitemOpen
  \bibfield  {author} {\bibinfo {author} {\bibfnamefont {R.}~\bibnamefont
  {Vinuesa}}\ and\ \bibinfo {author} {\bibfnamefont {S.}~\bibnamefont
  {Brunton}},\ }\bibfield  {title} {\enquote {\bibinfo {title} {Emerging trends
  in machine learning for computational fluid dynamics},}\ }\href@noop {}
  {\bibfield  {journal} {\bibinfo  {journal} {arXiv preprint arXiv:2211.15145}\
  } (\bibinfo {year} {2022}{\natexlab{a}})}\BibitemShut {NoStop}%
\bibitem [{\citenamefont {Vinuesa}\ and\ \citenamefont
  {Brunton}(2022{\natexlab{b}})}]{vinuesa2022enhancing}%
  \BibitemOpen
  \bibfield  {author} {\bibinfo {author} {\bibfnamefont {R.}~\bibnamefont
  {Vinuesa}}\ and\ \bibinfo {author} {\bibfnamefont {S.~L.}\ \bibnamefont
  {Brunton}},\ }\bibfield  {title} {\enquote {\bibinfo {title} {Enhancing
  computational fluid dynamics with machine learning},}\ }\href@noop {}
  {\bibfield  {journal} {\bibinfo  {journal} {Nature Computational Science}\
  }\textbf {\bibinfo {volume} {2}},\ \bibinfo {pages} {358--366} (\bibinfo
  {year} {2022}{\natexlab{b}})}\BibitemShut {NoStop}%
\bibitem [{\citenamefont {Erichson}\ \emph {et~al.}(2020)\citenamefont
  {Erichson}, \citenamefont {Mathelin}, \citenamefont {Yao}, \citenamefont
  {Brunton}, \citenamefont {Mahoney},\ and\ \citenamefont
  {Kutz}}]{erichson2020shallow}%
  \BibitemOpen
  \bibfield  {author} {\bibinfo {author} {\bibfnamefont {N.~B.}\ \bibnamefont
  {Erichson}}, \bibinfo {author} {\bibfnamefont {L.}~\bibnamefont {Mathelin}},
  \bibinfo {author} {\bibfnamefont {Z.}~\bibnamefont {Yao}}, \bibinfo {author}
  {\bibfnamefont {S.~L.}\ \bibnamefont {Brunton}}, \bibinfo {author}
  {\bibfnamefont {M.~W.}\ \bibnamefont {Mahoney}}, \ and\ \bibinfo {author}
  {\bibfnamefont {J.~N.}\ \bibnamefont {Kutz}},\ }\bibfield  {title} {\enquote
  {\bibinfo {title} {Shallow neural networks for fluid flow reconstruction with
  limited sensors},}\ }\href@noop {} {\bibfield  {journal} {\bibinfo  {journal}
  {Proceedings of the Royal Society A}\ }\textbf {\bibinfo {volume} {476}},\
  \bibinfo {pages} {20200097} (\bibinfo {year} {2020})}\BibitemShut {NoStop}%
\bibitem [{\citenamefont {Donoho}(2006)}]{donoho2006compressed}%
  \BibitemOpen
  \bibfield  {author} {\bibinfo {author} {\bibfnamefont {D.~L.}\ \bibnamefont
  {Donoho}},\ }\bibfield  {title} {\enquote {\bibinfo {title} {Compressed
  sensing},}\ }\href@noop {} {\bibfield  {journal} {\bibinfo  {journal} {IEEE
  Transactions on information theory}\ }\textbf {\bibinfo {volume} {52}},\
  \bibinfo {pages} {1289--1306} (\bibinfo {year} {2006})}\BibitemShut {NoStop}%
\bibitem [{\citenamefont {Deng}\ \emph {et~al.}(2019)\citenamefont {Deng},
  \citenamefont {He}, \citenamefont {Liu},\ and\ \citenamefont
  {Kim}}]{deng2019super}%
  \BibitemOpen
  \bibfield  {author} {\bibinfo {author} {\bibfnamefont {Z.}~\bibnamefont
  {Deng}}, \bibinfo {author} {\bibfnamefont {C.}~\bibnamefont {He}}, \bibinfo
  {author} {\bibfnamefont {Y.}~\bibnamefont {Liu}}, \ and\ \bibinfo {author}
  {\bibfnamefont {K.~C.}\ \bibnamefont {Kim}},\ }\bibfield  {title} {\enquote
  {\bibinfo {title} {Super-resolution reconstruction of turbulent velocity
  fields using a generative adversarial network-based artificial intelligence
  framework},}\ }\href@noop {} {\bibfield  {journal} {\bibinfo  {journal}
  {Physics of Fluids}\ }\textbf {\bibinfo {volume} {31}},\ \bibinfo {pages}
  {125111} (\bibinfo {year} {2019})}\BibitemShut {NoStop}%
\bibitem [{\citenamefont {Fukami}, \citenamefont {Fukagata},\ and\
  \citenamefont {Taira}(2019)}]{fukami2019super}%
  \BibitemOpen
  \bibfield  {author} {\bibinfo {author} {\bibfnamefont {K.}~\bibnamefont
  {Fukami}}, \bibinfo {author} {\bibfnamefont {K.}~\bibnamefont {Fukagata}}, \
  and\ \bibinfo {author} {\bibfnamefont {K.}~\bibnamefont {Taira}},\ }\bibfield
   {title} {\enquote {\bibinfo {title} {Super-resolution reconstruction of
  turbulent flows with machine learning},}\ }\href@noop {} {\bibfield
  {journal} {\bibinfo  {journal} {Journal of Fluid Mechanics}\ }\textbf
  {\bibinfo {volume} {870}},\ \bibinfo {pages} {106--120} (\bibinfo {year}
  {2019})}\BibitemShut {NoStop}%
\bibitem [{\citenamefont {Liu}\ \emph {et~al.}(2020)\citenamefont {Liu},
  \citenamefont {Tang}, \citenamefont {Huang},\ and\ \citenamefont
  {Lu}}]{liu2020deep}%
  \BibitemOpen
  \bibfield  {author} {\bibinfo {author} {\bibfnamefont {B.}~\bibnamefont
  {Liu}}, \bibinfo {author} {\bibfnamefont {J.}~\bibnamefont {Tang}}, \bibinfo
  {author} {\bibfnamefont {H.}~\bibnamefont {Huang}}, \ and\ \bibinfo {author}
  {\bibfnamefont {X.-Y.}\ \bibnamefont {Lu}},\ }\bibfield  {title} {\enquote
  {\bibinfo {title} {Deep learning methods for super-resolution reconstruction
  of turbulent flows},}\ }\href@noop {} {\bibfield  {journal} {\bibinfo
  {journal} {Physics of Fluids}\ }\textbf {\bibinfo {volume} {32}},\ \bibinfo
  {pages} {025105} (\bibinfo {year} {2020})}\BibitemShut {NoStop}%
\bibitem [{\citenamefont {Gao}, \citenamefont {Sun},\ and\ \citenamefont
  {Wang}(2021)}]{gao2021super}%
  \BibitemOpen
  \bibfield  {author} {\bibinfo {author} {\bibfnamefont {H.}~\bibnamefont
  {Gao}}, \bibinfo {author} {\bibfnamefont {L.}~\bibnamefont {Sun}}, \ and\
  \bibinfo {author} {\bibfnamefont {J.-X.}\ \bibnamefont {Wang}},\ }\bibfield
  {title} {\enquote {\bibinfo {title} {Super-resolution and denoising of fluid
  flow using physics-informed convolutional neural networks without
  high-resolution labels},}\ }\href@noop {} {\bibfield  {journal} {\bibinfo
  {journal} {Physics of Fluids}\ }\textbf {\bibinfo {volume} {33}},\ \bibinfo
  {pages} {073603} (\bibinfo {year} {2021})}\BibitemShut {NoStop}%
\bibitem [{\citenamefont {Yousif}, \citenamefont {Yu},\ and\ \citenamefont
  {Lim}(2021)}]{yousif2021high}%
  \BibitemOpen
  \bibfield  {author} {\bibinfo {author} {\bibfnamefont {M.~Z.}\ \bibnamefont
  {Yousif}}, \bibinfo {author} {\bibfnamefont {L.}~\bibnamefont {Yu}}, \ and\
  \bibinfo {author} {\bibfnamefont {H.-C.}\ \bibnamefont {Lim}},\ }\bibfield
  {title} {\enquote {\bibinfo {title} {High-fidelity reconstruction of
  turbulent flow from spatially limited data using enhanced super-resolution
  generative adversarial network},}\ }\href@noop {} {\bibfield  {journal}
  {\bibinfo  {journal} {Physics of Fluids}\ }\textbf {\bibinfo {volume} {33}},\
  \bibinfo {pages} {125119} (\bibinfo {year} {2021})}\BibitemShut {NoStop}%
\bibitem [{\citenamefont {Matsuo}\ \emph {et~al.}(2021)\citenamefont {Matsuo},
  \citenamefont {Nakamura}, \citenamefont {Morimoto}, \citenamefont {Fukami},\
  and\ \citenamefont {Fukagata}}]{matsuo2021supervised}%
  \BibitemOpen
  \bibfield  {author} {\bibinfo {author} {\bibfnamefont {M.}~\bibnamefont
  {Matsuo}}, \bibinfo {author} {\bibfnamefont {T.}~\bibnamefont {Nakamura}},
  \bibinfo {author} {\bibfnamefont {M.}~\bibnamefont {Morimoto}}, \bibinfo
  {author} {\bibfnamefont {K.}~\bibnamefont {Fukami}}, \ and\ \bibinfo {author}
  {\bibfnamefont {K.}~\bibnamefont {Fukagata}},\ }\bibfield  {title} {\enquote
  {\bibinfo {title} {Supervised convolutional network for three-dimensional
  fluid data reconstruction from sectional flow fields with adaptive
  super-resolution assistance},}\ }\href@noop {} {\bibfield  {journal}
  {\bibinfo  {journal} {arXiv preprint arXiv:2103.09020}\ } (\bibinfo {year}
  {2021})}\BibitemShut {NoStop}%
\bibitem [{\citenamefont {G{\"u}emes}\ \emph {et~al.}(2021)\citenamefont
  {G{\"u}emes}, \citenamefont {Discetti}, \citenamefont {Ianiro}, \citenamefont
  {Sirmacek}, \citenamefont {Azizpour},\ and\ \citenamefont
  {Vinuesa}}]{guemes2021coarse}%
  \BibitemOpen
  \bibfield  {author} {\bibinfo {author} {\bibfnamefont {A.}~\bibnamefont
  {G{\"u}emes}}, \bibinfo {author} {\bibfnamefont {S.}~\bibnamefont
  {Discetti}}, \bibinfo {author} {\bibfnamefont {A.}~\bibnamefont {Ianiro}},
  \bibinfo {author} {\bibfnamefont {B.}~\bibnamefont {Sirmacek}}, \bibinfo
  {author} {\bibfnamefont {H.}~\bibnamefont {Azizpour}}, \ and\ \bibinfo
  {author} {\bibfnamefont {R.}~\bibnamefont {Vinuesa}},\ }\bibfield  {title}
  {\enquote {\bibinfo {title} {From coarse wall measurements to turbulent
  velocity fields through deep learning},}\ }\href@noop {} {\bibfield
  {journal} {\bibinfo  {journal} {Physics of Fluids}\ }\textbf {\bibinfo
  {volume} {33}},\ \bibinfo {pages} {075121} (\bibinfo {year}
  {2021})}\BibitemShut {NoStop}%
\bibitem [{\citenamefont {Zhou}\ \emph {et~al.}(2022)\citenamefont {Zhou},
  \citenamefont {Li}, \citenamefont {Yang},\ and\ \citenamefont
  {Yang}}]{zhou2022robust}%
  \BibitemOpen
  \bibfield  {author} {\bibinfo {author} {\bibfnamefont {Z.}~\bibnamefont
  {Zhou}}, \bibinfo {author} {\bibfnamefont {B.}~\bibnamefont {Li}}, \bibinfo
  {author} {\bibfnamefont {X.}~\bibnamefont {Yang}}, \ and\ \bibinfo {author}
  {\bibfnamefont {Z.}~\bibnamefont {Yang}},\ }\bibfield  {title} {\enquote
  {\bibinfo {title} {A robust super-resolution reconstruction model of
  turbulent flow data based on deep learning},}\ }\href@noop {} {\bibfield
  {journal} {\bibinfo  {journal} {Computers \& Fluids}\ }\textbf {\bibinfo
  {volume} {239}},\ \bibinfo {pages} {105382} (\bibinfo {year}
  {2022})}\BibitemShut {NoStop}%
\bibitem [{\citenamefont {Yousif}, \citenamefont {Yu},\ and\ \citenamefont
  {Lim}(2022)}]{yousif2022super}%
  \BibitemOpen
  \bibfield  {author} {\bibinfo {author} {\bibfnamefont {M.~Z.}\ \bibnamefont
  {Yousif}}, \bibinfo {author} {\bibfnamefont {L.}~\bibnamefont {Yu}}, \ and\
  \bibinfo {author} {\bibfnamefont {H.-C.}\ \bibnamefont {Lim}},\ }\bibfield
  {title} {\enquote {\bibinfo {title} {Super-resolution reconstruction of
  turbulent flow fields at various reynolds numbers based on generative
  adversarial networks},}\ }\href@noop {} {\bibfield  {journal} {\bibinfo
  {journal} {Physics of Fluids}\ }\textbf {\bibinfo {volume} {34}},\ \bibinfo
  {pages} {015130} (\bibinfo {year} {2022})}\BibitemShut {NoStop}%
\bibitem [{\citenamefont {Yu}\ \emph {et~al.}(2022)\citenamefont {Yu},
  \citenamefont {Yousif}, \citenamefont {Zhang}, \citenamefont {Hoyas},
  \citenamefont {Vinuesa},\ and\ \citenamefont {Lim}}]{yu2022three}%
  \BibitemOpen
  \bibfield  {author} {\bibinfo {author} {\bibfnamefont {L.}~\bibnamefont
  {Yu}}, \bibinfo {author} {\bibfnamefont {M.~Z.}\ \bibnamefont {Yousif}},
  \bibinfo {author} {\bibfnamefont {M.}~\bibnamefont {Zhang}}, \bibinfo
  {author} {\bibfnamefont {S.}~\bibnamefont {Hoyas}}, \bibinfo {author}
  {\bibfnamefont {R.}~\bibnamefont {Vinuesa}}, \ and\ \bibinfo {author}
  {\bibfnamefont {H.-C.}\ \bibnamefont {Lim}},\ }\bibfield  {title} {\enquote
  {\bibinfo {title} {Three-dimensional esrgan for super-resolution
  reconstruction of turbulent flows with tricubic interpolation-based transfer
  learning},}\ }\href@noop {} {\bibfield  {journal} {\bibinfo  {journal}
  {Physics of Fluids}\ }\textbf {\bibinfo {volume} {34}},\ \bibinfo {pages}
  {125126} (\bibinfo {year} {2022})}\BibitemShut {NoStop}%
\bibitem [{\citenamefont {Zhang}\ \emph {et~al.}(2022)\citenamefont {Zhang},
  \citenamefont {Xiao}, \citenamefont {Choi},\ and\ \citenamefont
  {Mao}}]{zhang2022fusion}%
  \BibitemOpen
  \bibfield  {author} {\bibinfo {author} {\bibfnamefont {Z.}~\bibnamefont
  {Zhang}}, \bibinfo {author} {\bibfnamefont {D.}~\bibnamefont {Xiao}},
  \bibinfo {author} {\bibfnamefont {K.-S.}\ \bibnamefont {Choi}}, \ and\
  \bibinfo {author} {\bibfnamefont {X.}~\bibnamefont {Mao}},\ }\bibfield
  {title} {\enquote {\bibinfo {title} {The fusion of flow field data with
  multiple fidelities},}\ }\href@noop {} {\bibfield  {journal} {\bibinfo
  {journal} {Physics of Fluids}\ }\textbf {\bibinfo {volume} {34}},\ \bibinfo
  {pages} {097113} (\bibinfo {year} {2022})}\BibitemShut {NoStop}%
\bibitem [{\citenamefont {Mondal}\ and\ \citenamefont
  {Sarkar}(2022)}]{mondal2022multi}%
  \BibitemOpen
  \bibfield  {author} {\bibinfo {author} {\bibfnamefont {S.}~\bibnamefont
  {Mondal}}\ and\ \bibinfo {author} {\bibfnamefont {S.}~\bibnamefont
  {Sarkar}},\ }\bibfield  {title} {\enquote {\bibinfo {title} {Multi-fidelity
  prediction of spatiotemporal fluid flow},}\ }\href@noop {} {\bibfield
  {journal} {\bibinfo  {journal} {Physics of Fluids}\ }\textbf {\bibinfo
  {volume} {34}},\ \bibinfo {pages} {087112} (\bibinfo {year}
  {2022})}\BibitemShut {NoStop}%
\bibitem [{\citenamefont {Samtaney}, \citenamefont {Pullin},\ and\
  \citenamefont {Kosovi{\'c}}(2001)}]{samtaney2001direct}%
  \BibitemOpen
  \bibfield  {author} {\bibinfo {author} {\bibfnamefont {R.}~\bibnamefont
  {Samtaney}}, \bibinfo {author} {\bibfnamefont {D.~I.}\ \bibnamefont
  {Pullin}}, \ and\ \bibinfo {author} {\bibfnamefont {B.}~\bibnamefont
  {Kosovi{\'c}}},\ }\bibfield  {title} {\enquote {\bibinfo {title} {Direct
  numerical simulation of decaying compressible turbulence and shocklet
  statistics},}\ }\href@noop {} {\bibfield  {journal} {\bibinfo  {journal}
  {Physics of Fluids}\ }\textbf {\bibinfo {volume} {13}},\ \bibinfo {pages}
  {1415--1430} (\bibinfo {year} {2001})}\BibitemShut {NoStop}%
\bibitem [{\citenamefont {Johnsen}\ \emph {et~al.}(2010)\citenamefont
  {Johnsen}, \citenamefont {Larsson}, \citenamefont {Bhagatwala}, \citenamefont
  {Cabot}, \citenamefont {Moin}, \citenamefont {Olson}, \citenamefont {Rawat},
  \citenamefont {Shankar}, \citenamefont {Sj{\"o}green}, \citenamefont {Yee}
  \emph {et~al.}}]{johnsen2010assessment}%
  \BibitemOpen
  \bibfield  {author} {\bibinfo {author} {\bibfnamefont {E.}~\bibnamefont
  {Johnsen}}, \bibinfo {author} {\bibfnamefont {J.}~\bibnamefont {Larsson}},
  \bibinfo {author} {\bibfnamefont {A.~V.}\ \bibnamefont {Bhagatwala}},
  \bibinfo {author} {\bibfnamefont {W.~H.}\ \bibnamefont {Cabot}}, \bibinfo
  {author} {\bibfnamefont {P.}~\bibnamefont {Moin}}, \bibinfo {author}
  {\bibfnamefont {B.~J.}\ \bibnamefont {Olson}}, \bibinfo {author}
  {\bibfnamefont {P.~S.}\ \bibnamefont {Rawat}}, \bibinfo {author}
  {\bibfnamefont {S.~K.}\ \bibnamefont {Shankar}}, \bibinfo {author}
  {\bibfnamefont {B.}~\bibnamefont {Sj{\"o}green}}, \bibinfo {author}
  {\bibfnamefont {H.~C.}\ \bibnamefont {Yee}},  \emph {et~al.},\ }\bibfield
  {title} {\enquote {\bibinfo {title} {Assessment of high-resolution methods
  for numerical simulations of compressible turbulence with shock waves},}\
  }\href@noop {} {\bibfield  {journal} {\bibinfo  {journal} {Journal of
  Computational Physics}\ }\textbf {\bibinfo {volume} {229}},\ \bibinfo {pages}
  {1213--1237} (\bibinfo {year} {2010})}\BibitemShut {NoStop}%
\bibitem [{\citenamefont {Saadat}\ \emph {et~al.}(2021)\citenamefont {Saadat},
  \citenamefont {Hosseini}, \citenamefont {Dorschner},\ and\ \citenamefont
  {Karlin}}]{saadat2021extended}%
  \BibitemOpen
  \bibfield  {author} {\bibinfo {author} {\bibfnamefont {M.~H.}\ \bibnamefont
  {Saadat}}, \bibinfo {author} {\bibfnamefont {S.~A.}\ \bibnamefont
  {Hosseini}}, \bibinfo {author} {\bibfnamefont {B.}~\bibnamefont {Dorschner}},
  \ and\ \bibinfo {author} {\bibfnamefont {I.}~\bibnamefont {Karlin}},\
  }\bibfield  {title} {\enquote {\bibinfo {title} {Extended lattice boltzmann
  model for gas dynamics},}\ }\href@noop {} {\bibfield  {journal} {\bibinfo
  {journal} {Physics of Fluids}\ }\textbf {\bibinfo {volume} {33}},\ \bibinfo
  {pages} {046104} (\bibinfo {year} {2021})}\BibitemShut {NoStop}%
\bibitem [{\citenamefont {Saadat}, \citenamefont {B{\"o}sch},\ and\
  \citenamefont {Karlin}(2019)}]{saadat2019lattice}%
  \BibitemOpen
  \bibfield  {author} {\bibinfo {author} {\bibfnamefont {M.~H.}\ \bibnamefont
  {Saadat}}, \bibinfo {author} {\bibfnamefont {F.}~\bibnamefont {B{\"o}sch}}, \
  and\ \bibinfo {author} {\bibfnamefont {I.~V.}\ \bibnamefont {Karlin}},\
  }\bibfield  {title} {\enquote {\bibinfo {title} {Lattice boltzmann model for
  compressible flows on standard lattices: Variable prandtl number and
  adiabatic exponent},}\ }\href@noop {} {\bibfield  {journal} {\bibinfo
  {journal} {Physical Review E}\ }\textbf {\bibinfo {volume} {99}},\ \bibinfo
  {pages} {013306} (\bibinfo {year} {2019})}\BibitemShut {NoStop}%
\end{thebibliography}%

\end{document}